\documentclass[aps,prl,twocolumn,groupedaddress]{revtex4}
\usepackage[section]{placeins}        %Keep Figure in Section
\usepackage{float}% If comment this, figure moves to Page 2
\usepackage{kpfonts}
 \usepackage{graphics}
 \usepackage{multirow}
\usepackage{epsfig}
\usepackage{pdfpages}
\usepackage{amsmath,amsthm, amssymb, latexsym}
\usepackage{color}
\usepackage{graphics}
\usepackage{epsfig}
\usepackage[outercaption]{sidecap}%\usepackage{ulem}

\newcommand{\bee}{\begin{equation}}
\newcommand{\ene}{\end{equation}}
\newcommand{\beea}{\begin{eqnarray}}
\newcommand{\enea}{\end{eqnarray}}

\begin{document}
\title{Fast Ignition Laser Fusion Using In-Situ Ion Acceleration With Pulsed CO$ _2 $ Lasers }
%\author{abc}
\author{Atul Kumar}
\email{atul.j1211@gmail.com}
\author{Chandrasekar Shukla}
%%\author{Amita Das} 
\author{Predhiman Kaw$^0$}
\author{Amita Das}
\email{amita@ipr.res.in}
\affiliation{Institute for Plasma Research, HBNI, Bhat, Gandhinagar - 382428, India }
%\date{\today}
%{for revtex  maketitle should be written just here}
%\maketitle 
\begin{abstract} 
Fast ignition  is an alternative concept of laser fusion in which the task of compressing the fusion pellet to supersolid densities is accomplished by the conventional high energy nanosecond glass lasers and the task of igniting the compressed pellet is given to a high intensity, moderate energy pico second source which can set the pellet ablaze  (with DT fusion reactions) by creating an adequate size hot spot in it. In this letter, we present a conceptual method by which energy from carbon dioxide ($ CO_2 $) laser could be coupled to heat up ions  produced in situ in the plasma, which in turn produces the required hot spot.  An efficient conversion of $ CO_2 $ energy into ion beam energy can thus give us the required source of energy for the fast ignition laser fusion. We demonstrate by using PIC (Particle - In - Cell) simulations that the use of  several Kilo Tesla of an external magnetic field in the transverse direction in an inhomogeneous plasma where a cutoff is followed by a resonance can lead to  coupling of laser energy to a lower hybrid ion plasma resonance. Then ions are accelerated efficiently ($\sim 1 MeV$) by the breaking of lower hybrid waves and thereby enabling the possibility of  fast ignition by ions  in a convenient fashion.
\end{abstract}
% for revtex4 here maketitle should be written
%\pacs{52.30.Cv,52.35.Ra} 
\maketitle 
%\begin{multicols}{2}
% upto here ------

Fast ignition \cite{Tabak,Kemp2014} is an alternative concept of laser fusion \cite{RevModPhys.46.325,Betti2016} in which the task of compressing the fusion pellet to super solid densities $ (\sim 1000 gm/cc) $ or more is accompolished by conventional high energy nanosecond glass lasers and the task of igniting the  compressed pellet is given to a seperate high intensity, moderate energy pico second energy source which can set pellet ablaze (with DT fusion reactions) by creating an adequate size hot spot with ignition temperatures $ \sim 5 KeV $ in it. Many alternatives have been considered for initiating the fast ignition heating process. The use of very high intensity $ 1 $ micron laser picosecond pulses have not proven to be efficacious because the laser energy ends up in a relativistic electron beam at overdense compressed surface, which creates massive forward and return currents, current instabilities and filamentations of the beam on a rapid electron time scale \cite{Beg1997,Wilks1992,Kumar2017,Weibel,Buneman1959,Bell1964,Roberts1967,Bret2009,Bret2010,EMHD,cshukla}. These instabilities lead to a divergence of the incident electron beam away from the central compressed target, thus preventing the formation of a hot spot in the interior. Other schemes considered, include mitigation of instabilities leading to turbulent magnetic fields  with externally produced guiding magnetic fields ( in the range of $ 1-10 Kilo-Tesla $) \cite{Sheng_96,Knauer_2010,Yoneda_2012,Nagatomo2013,Fujioka2016,Matsuo2017,Sakata2017}, initiation using shock waves,  using externally produced ion beams etc., but none so far have shown  significantly improved results.\footnotetext[0]{Predhiman Kaw is a deceased author.}

The $ CO_2 $ lasers are today the most efficient industrial lasers in the market and immediately come to mind where continuous wave (CW) and relatively long pulse applications are concerned \cite{Haberberger2010,Tochitsky2016}. They have not been as widely explored for the short pulse (sub-nano or pico second)  applications. It is, therefore, not as widely known that chirped pulse amplification (CPA) techniques \cite{Strickland1985} have been applied to $ CO_2 $ lasers. Right now $ 100 TW $, $ 1 psec $ lasers with an energy content of about 100 joules are on anvil (after successful demonstration of $ 10 TW $, $ 1 psec $ pulses). An energy range of about $ \sim 10-100 KJ $, $ \sim 10 psec $ pulses would be of direct interest to fast ignition fusion if a method could be found of delivering about $ 10 \% $ of this energy to a hot spot in the compressed matter. 
 
In this letter, we present a conceptual study followed up by the PIC simulations through OSIRIS-4.0 \cite{hemker,Fonseca2002,osiris},  of a method by which energy from the $ CO_2 $ laser could be coupled through lower hybrid excitation to plasma  for accelerating ions.  
These accelerated ions could, in turn, produce the required hot spot. 
%Recently, a right-hand circularly polarized electromagnetic wave has been shown to propagate in a strongly magnetized plasma of any density
% without encountering cutoff in the whistler mode \cite{Luan_2016} which leads to plasma heating up to $10 KeV$ via electron-ion collisions. 
If we consider ion heating for fast ignition of a fusion pellet, the requirements may be put as ion energy $ \sim 10 MeV $, beam current $ ~250 MAmps $, beam period $ ~10 psec $ which is a $ 5 KJ $ beam pulse. An efficient conversion of $ CO_2$ energy into ion beam energy can thus give the required source of energy for fast ignition laser fusion. The detailed mechanism is derived below. The multiple advantages of using in situ accelerated ions as  projectiles for delievering the energy to fusion fuel in the hot spot are: (1) Efficient coupling directly to fusion species, viz. background DT species ; (2) Ions being non relativistic and total current being much less than Alfven critical current for ions, magnetic fluctuations are expected to play negligible role in beam transport and beam stopping; (3) Classical collisional stopping of few MeV ions in dense core will be adequate; (4) Return currents, if any, will excite electrostatic ion acoustic modes leading to anomalous heating of the background plasma.

How do we utilize the $ CO_2 $ laser for giving energy preferentially to ions? One point of departure from the past experience could be to use magnetized plasma modes for the interaction of the $ CO_2 $ laser with the plasma. Now this is feasible since the external fields of the order $ 10^4 $ Tesla have already been used in the laboratory \cite{Bailly-Grandvaux2018}. As the wavelength of this laser is $ 10 $
times that of a glass laser, at a plasma density of $ 10^{19} cm^{-3} $, we have $ \omega \sim \omega_{pe} \ll \omega_{ce} \sim	2 \times 10^{15} rad\cdot sec^{-1} $ at a magnetic field of $10^4$ Tesla. Thus the magnetic field very strongly modifies the propagation physics of the laser beam through the plasma. 
In particular, if we choose X-mode of propagation for the entry of the laser into the plasma, the first cut-off (viz. right handed cutoff) and the first resonance (viz. the upper hybrid resonance) will not be encountered by the laser as it enters the plasma from the low density side. Similarly, the density for a corresponding resonance required for nonlinear Raman scattering \cite{Joshi1981,Rozmus1987,Mori1994} of perpendicularly propagating electron modes will also be excluded from plasma profile.  
%{\textcolor{red}{(Supplementary materials provide details for these estimates)}}. 
It will thus become possible to use the second cut-off (the left handed cut-off) and the second resonance, namely the lower hybrid resonance, for the interaction of the intense $ CO_2 $ laser light with the plasma. The laser propagation and the density profile 
has been illustrated  in the schematic plot of Fig. 1(a).  We   use transverse magnetic field at the overdense laser plasma interface to immobilize the electrons due to strong magnetization and let ions  as an unmagnetized species be accelerated by short wavelength (much shorter than ion Larmor radius) electrostatic waves such as lower hybrid resonance. In the linear regime, this suggests the utilization of the left handed cutoff and the lower hybrid resonance for the X-mode propagation \cite{Stix1960,Stix1965,Golant1972,P.BellanandM.Porkolab1975,Brambilla1976}. 
%If the density gradient on the surface is very sharp, one may have to use nonlinear mechanisms such as Brunnel like mechanisms \cite{Brunel1987} operating around the lower hybrid frequency. Noting that this frequency is of the order of ion plasma frequency 
%%($ \omega_{pe} \ll \omega_{ce} $), 
%the critical resonance layer now operates at relatively high density. In the case of $ CO_2 $ lasers this density is of the order of $ 3.2 \times 10^{20} cm^{-3} $  

  Now, consider a  propagation of electromagnetic waves in the X-mode configuration and the density gradient scale lengths 
   to be  long enough such that the resonances and cutoffs are well separated  from each other and identifiable. It is well known that an electromagnetic wave propagating in the X-mode typically couples with the longitudinal modes at the upper and lower hybrid frequencies, giving interesting absorption phenomena \cite{Budden1962,White1974}. 
   However, as stated earlier, the upper hybrid resonance has been ruled out in the present case due to the choice of the laser frequency in 
  comparison to the electron cyclotron frequency. 
  The dispersion relation for X-mode including ion response can be rewritten as 
\beea
\frac{c^2 k^2}{\omega^2} = \frac{\epsilon_L \cdot \epsilon_R}{\epsilon_L + \epsilon_R}
\label{disprelX}
\enea
where,
\begin{eqnarray}
\epsilon_R &=&1-\frac{\omega_{pe}^2}{\omega (\omega	 -\omega_{ce})} - \frac{\omega_{pi}^2}{\omega (\omega	 +\omega_{ci})}\\
\epsilon_L &=&1-\frac{\omega_{pe}^2}{\omega (\omega	 +\omega_{ce})} - \frac{\omega_{pi}^2}{\omega (\omega	 -\omega_{ci})}
\label{epsLepsR}
\end{eqnarray}
An examination of the resonances and cut offs show that for the choice of $ \omega \ll \omega_{ce} $, the first cutoff that is encountered by the incoming laser beam is the left handed cut off $ \epsilon_L = 0 $ given in  Eq.(\ref{epsLepsR}). It has been shown by  Budden \cite{Budden1962} and others that when an electromagnetic (EM) wave is incident from the low density side with cut off followed by resonance, the sum of the reflection and transmission coefficients does not add up to unity. This puzzle was solved by Budden \cite{Budden1962} when he showed by a mathematical analysis that the difference of energy is absorbed because of the  coupling to electrostatic waves in the resonance region. In our situation, the left handed cut off is followed by the lower hybrid resonance. The absorption process of this resonance will thus put energy into lower hybrid waves. When the cut off and the resonance are close to each other, we can write the wave equation near these two points in the following form: 
  \beea
\frac{d^2\vec{E}}{dx^2} + \frac{\omega^2}{c^2}\left(1+ \frac{x_0}{x}\right)\vec{E} =0
\label{BuddenEq}
\enea
where it is assumed that $ x  $ is the direction of the density gradient (as shown in Fig. 1(a)), magnetic field is pointing in the $ y - z $ plane  and the dominant component of the laser electric field of the X-mode is in the $ y $ direction. Eq.(\ref{BuddenEq}) shows a resonance at $ x=0 $ and the cut off at $ x= -x_0 $. The solution of  Eq.(\ref{BuddenEq}) can be written in terms of the Whittaker functions. Carrying out the standard asymtotics to investigate the situations when the waves are incident on the cut off region from the low density side and get partially reflected and partially transmitted one finds that,
\begin{eqnarray}
\vert T \vert = e^{-\pi s_0 /2}; \vert R \vert =1- e^{-\pi s_0 }
\label{RTcoeff}
\end{eqnarray}
where, $ s_0 = \frac{\omega}{c} x_0 $. we can note that squares of $ \vert T \vert $ and $ \vert R \vert $ do not add up to unity and that energy absorbed by the resonance is  of the order of 
\begin{eqnarray}
1 -{\vert T \vert}^2 - {\vert R \vert }^2 = (e^{-\pi s_0 } - e^{-2\pi s_0 })
\label{AbsEne}
\end{eqnarray}
This fractional absorption maximizes at $ \pi {s_0}= \ln2 $ and has a value  $ 1/4 $. This $ 25 \% $ of the incident laser energy is absorbed for the optimum case. Note that the above expression is based on the asymptotics and may only give an order of magnitude estimate and is not strictly accurate at the low values of $s_0$. To determine $ s_0 $, we need to find $ x_0 $ namely the distance between the reonance and the cut off. We restrict our attention to frequencies satisfying $ \omega_{ci} \ll\omega \ll \omega_{ce} $. The densities at the left handed cut off and the neighbouring lower hybrid resonance point  are thus given by the following expressions respectively:  
\begin{eqnarray}
\frac{1}{\omega_{pi_1}^2} = \frac{1}{\omega^2}\bigg(1+\frac{\omega}{\omega_{ci}}\bigg)\\
\frac{1}{\omega_{pi_2}^2} = \frac{1}{\omega^2}\bigg(1-\frac{\omega^2}{\omega_{ce}\omega_{ci}}\bigg)
\label{LCfreq}
\end{eqnarray}
It may be noted  from the expression of $ \omega_{pi_2} $ that $ \omega^2 $ is chosen below $ \omega_{ci} \omega_{ce} $. 
%This means that the magnetic field and the laser frequency have to satisfy more stringent condition of 
%\beea
%\frac{\omega^2}{\omega_{ce}^2} < \frac{m_e}{m_i}
%\label{sqrtmemi}
%\enea 
%For $ CO_2 $ laser frequencies, this correspond to a magnetic field of $ 14.13 $ kilo Tesla. 
Assuming a linear density profile between the cut off and the resonance point, we have 
\beea
\omega_{pi_2}^2 - \omega_{pi_1}^2 = x_0 \frac{d}{dx} \omega_{pi}^2
\label{difference}
\enea
For a linear density profile between the cut off and the resonance points, we have $ s_0 = \frac{\omega}{c} x_0 $.Thus the  expression of $s_0$ is given as,
\beea
s_0 = \frac{\omega}{c}\bigg(\frac{\omega^2}{\frac{d}{dx}\omega_{pi}^2}\bigg) \bigg[\bigg(1-\frac{\omega^2}{\omega_{ce}\omega_{ci}}\bigg)^{-1}- \bigg(1+\frac{\omega}{\omega_{ci}}\bigg)^{-1}\bigg]
\label{difference}
\enea
Note that the optimal parameter $ s_0 = \frac{\ln2}{\pi} $. We can then have the optimal scale length for density gradient length $ L_n = {\omega^2}/({\frac{d}{dx}\omega_{pi}^2})$. 
%$=1.611 \lambda_L = 16.11 \mu m $ where $\lambda_L = 10 \mu m$ is the  wavelength of the $CO_2$ laser.

Next we  make an estimate of the magnitude of ion acceleration during the lower hybrid resonance. Detailed calculations by Budden \cite{Budden1962} have shown that the energy absorbed in the above resonance absorption process is picked up by electrostatic lower hybrid waves which are dissipated either by collisions or by wave particle interactions. If the intensity of the wave is strong enough, this can lead to wave breaking and energy will be primarily picked up by the ions. Condition of wave breaking is approximately given by $ \frac{keE}{m_i \omega ^2} > 1 $. This implies that the velocity excursion of the ion in the wave field exceeds the phase velocity of the wave so that even the cold background ions are brought into resonance with the waves accelerated by it. The condition may also be written in the form of $ e \phi	=\frac{eE}{k} = m_i \frac{\omega^2}{k^2} $, that is the kinetic energy picked up by the ions is of the order of the wave potential itself. {We now estimate the magnitude  electric field of the wave that will be excited in the cut off resonance process. Taking  $\hat{x}$ as the direction of propagation of the perturbed mode, $ \vec{E} = E (\hat{x} + \alpha\hat{y}) e^{\iota(k\cdot x - \omega t)} $ where $E$ represents  amplitude of the electric field.}

\beea
\alpha = \frac{E_y}{E_x} = \frac{(\frac{\omega_{ce}}{\omega}) \frac{\omega_{pe}^2}{\omega_{ce}^2 - \omega^2} - (\frac{\omega_{ci}}{\omega}) \frac{\omega_{pi}^2}{\omega_{ci}^2 - \omega^2}}{\frac{(\omega^2 - \omega_{1}^2)(\omega^2 - \omega_{2}^2)}{\omega^2 (\omega^2 - \omega_{UH}^2)} - \bigg(1+ \frac{\omega_{pe}^2}{\omega_{ce}^2 - \omega^2} +\frac{\omega_{pi}^2}{\omega_{ci}^2 - \omega^2}\bigg)}
\label{sqrtmemi}
\enea 
where
\begin{eqnarray}
\omega_1= \frac{\omega_{ce}}{2}\bigg[ -1 + \bigg( 1+ \frac{4 \omega_{pe}^2}{\omega_{ce}^2}\bigg)^{1/2} \bigg]\\
\omega_2= \frac{\omega_{ce}}{2}\bigg[ 1 + \bigg( 1+ \frac{4 \omega_{pe}^2}{\omega_{ce}^2}\bigg)^{1/2} \bigg]\\
\mid \alpha \mid \rightarrow \sqrt{\frac{m_e}{m_i}} \frac{\omega_{pe}^2 \omega_{UH}^2}{\omega_{1}^2 \omega_{2}^2} \hspace{0.25cm}\forall \hspace{0.25cm} \omega \rightarrow \sqrt{\omega_{ce} \omega_{ci}}
\label{om1om2}
\end{eqnarray}

We now show with the help of PIC simulations   that the phenomena of lower hybrid excitation described above  indeed occurs
and is responsible for producing accelerated ions. 
For our simulation studies, we have chosen a  system of electrons and ions where ions are assumed to be $25$ times heavier than electrons (\emph{i.e.} $m_i = 25 m_e$ where $m_i$ and $m_e$ denotes the rest mass of the ion and electron species). For electrons to be strongly magnetized and ions to remain unmagnetized, we have considered  
$ \omega = \omega_{pi} = 2 \omega_{ci} $ and $ \omega_{ce} = 2.5 \omega_{pe} $ (i.e. around the density layer where ion plasma frequency is in resonance 
with laser frequency) . 
 Where $\omega_{ps} = \sqrt{4\pi n_0e^2/m_s}$ and $ \omega_{cs} = eB_0/m_sc$ are the plasma and cyclotron frequencies respectively for species $s $ 
which  corresponds to  electrons and ions. Here $e$ is the magnitude of the electronic charge, $n_0$ represents the critical density of the plasma and  $B_0 $ is the magnitude of the applied external magnetic field. The external magnetic field is   transverse to the density gradient, i.e. 
 $\vec{B} \cdot \vec{\nabla} n = 0$. 
The laser wave vector $ \vec{k}$ has a small  component parallel to the external magnetic field. 
The laser electric field has been chosen to fall obliquely at an angle of $\theta=60^{\circ}$ with normal to the plasma density surface so as to have  $ \vec{E} \cdot \vec{\nabla} n$ as finite.  
  These conditions are required for the excitation of  lower hybrid wave in the system. In order to accommodate an obliquely falling laser pulse 
  on constant density surfaces in simulations, we have oriented the 
  simulation box at an angle of $\theta = 60^{\circ}$ as shown in Fig. 1(b). 
  { In the new system,  we have the  laser propagation directed along $x^{\prime}$. The new coordinate axes have been represented by $x^{\prime} - y^{\prime}$ and they are related to the old system 
  by the transformation: \[
   \left[ {\begin{array}{cc}
   x^{\prime} \\
   y^{\prime} \\
  \end{array} } \right]=
  \left[ {\begin{array}{cc}
   \cos \theta & \sin \theta \\
   -\sin \theta & \cos \theta \\
  \end{array} } \right] \left[ {\begin{array}{cc}
   x \\
   y \\
  \end{array} } \right]
\]}

%For the system considered in this paper, the value of $\vert \alpha \vert = 1.17$ at the resonance point (\emph{i.e.} $ \omega \rightarrow \sqrt{\omega_{ce}\omega_{ci}}$).
%If the ion excursion length in the fields is of the order of the separation of the resonance and the cut off, the interface has to be regarded as a sharp interface and classical linear theory results absorption gives way for nonlinear brunnel type vacuum type heating mechanisms. this mechanism relies on the idea that the electrons are strongly magnetized and held by the magnetic fields where as the ion displacement perpendicular to a sharp interface (along the electric field) is much greater than the electron displacement which requires $ \omega^2 < {\omega_{ci} \omega_{ce}} $. Note that this satisfies our choice of laser frequency in the magnetic field. Under these conditions, ions come out in the vacuum whereas the electrons are firmly held back. The ions bunching in the vacuum is in response to the external field created by the laser such that at any given instance the field of the overdense interface is zero. That is the ion bunches shield the plasma from the normal component of the electric field. The motion of these ion bunches in the combine external and self consistent fields lead to an energization of the vacuum ions which re-enter the plasma in the form of  dense bunch or beams of the energetic ions. This may be seen from the expression below. To make further estimates of this effect, we have resorted to particle-in-cell (PIC) simulation with OSIRIS-4.0.

The numerical experiment is performed by using particle-in-cell (PIC) code OSIRIS-4.0 \cite{hemker,Fonseca2002,osiris}. The rectangular simulation box is $ 1050 c/\omega_{pe} $ in $x^{\prime}$-direction (the laser-propagation direction) and $ 500 c/\omega_{pe}$  in 
$y^{\prime}$-direction. The oblique target is composed of electron-ion plasma having  density profile linearly increasing from $0$ to $25 n_0$ where $n_0$ is $3.2 \times 10^{20} cm^{-3}$. The density has linear profile which can be expressed in a mathematical form as, $ n = \frac{x^{\prime}-\sqrt{3} y^{\prime}}{40} $. 
%The  ions have charge of proton and mass of an ion is $25 m_e$ where $m_e$ is the rest mass of the electron.
 A p-polarized plane $ CO_2 $ short pulse laser, having a wavelength of $10 \mu m$,  is launched from the left boundary. It strikes the plasma target   with an incident angle of $30^{\circ}$. The   peak intensities of $I = 7\times 10^{17} W/cm^2$ with a rise and fall time of $54.6217 \omega_{pe}^{-1}$ each has been chosen. The grid size is $0.1 c/\omega_{pe}$  in both direction with $ 10 $ cells per electron skin depth ($ c/\omega_{pe} $) and $ 64 $ particles per cell for each species. The charges are weighted according to the initial local density. The boundary conditions for the electromagnetic fields and particles are absorbing in both $x^{\prime}$ and $y^{\prime}$ direction. The time step  satisfying the  Courant condition is $ 0.07 \omega_{pe}^{-1} $. A  uniform external magnetic field of $B_{0z} =14.13$ kilo-Tesla is applied to strongly magnetize the electrons of the plasma along $\hat{z}$ direction. 

It is shown in Fig.~\ref{fig2} that the mode travels well beyond the lower hybrid resonance point which is around the line of density $ n \simeq 10 n_0 $ at $ t=1131.2 \omega_{pe}^{-1} $. The value of $\vert \alpha \vert = 1.25$ at this point and matches well with the analytically calculated value from Eq.(\ref{sqrtmemi}) which turns out to be $ 1.17 $.  This is a  clear indication of lower hybrid resonance. The  electric field energy spectra 
have been shown in Fig.~(\ref{fig3}) at various times. It should be noted that the spectra $k_{x^{\prime}}$ 
vs. $k_{y^{\prime}}$ plane remains  predominantly 
confined to  finite $k_{x^{\prime}}$ (the laser propagation wave vector) at early times. However, at the time of mode conversion $ t=1131.2 \omega_{pe}^{-1} $, power in finite $k_{y^{\prime}}$ starts acquiring power. 

We have also provided a comparison of  electron and ion  energy gain 
in the presence and absence of the external magnetic fields. A comparison of Fig(4) and Fig(5) wherein 
the electron and ion distribution in the  $p_{x^{\prime}}$ vs. $p_{y^{\prime}}$ with no external 
magnetic field shows that the energy acquired by the ions is quite small. The electron and ion counts 
for this case, plotted in Fig.(6) and Fig. (7) respectively, show that while electrons gain energy of the order of 
 $ \approx 8 MeV $, the ion energy in this case is relatively small ($ \approx 150 KeV $). 
 However, when the simulations are carried out in the presence of the external magnetic field, the energy  acquired by electrons has been reduced to $1MeV$ (Fig. (8)) and  the gain 
in ion energy is higher as expected by the lower hybrid resonance heating mechanism shown in Fig. (9). Both species (electrons and ions) are accelerated up to $ \sim 1MeV $ (Fig. (10-11)).  The approximate equipartition of electron and ion energies 
is also indicative of the lower hybrid resonance process. From the simulations, it is also evident  that the factor $\delta= {keE}/{m_i \omega ^2} = 1.34 $  which satisfies the breaking condition of the lower hybrid wave (\emph{i.e.} $ \delta > 1 $; where $k$ is the wavenumber of the mode, $E$ is the maximum electric field strength of the mode) near resonance region.

We have thus demonstrated the possibility of using lower hybrid resonance condition of  $CO_2$  lasers in the presence of external magnetic field for the in situ acceleration of plasma ions.  
When  pulsed CPA based high power  $CO_2$ lasers have become available and the magnetic field of the 
order of Kilo Tesla have been created in the laboratory recently, the mechanism of ion acceleration proposed in our work 
is ideally suited for experimentation.

\bibliographystyle{ieeetr}  

\bibliography{LH-heating}

\begin{figure*}[h!]
\center
                \includegraphics[width=\textwidth]{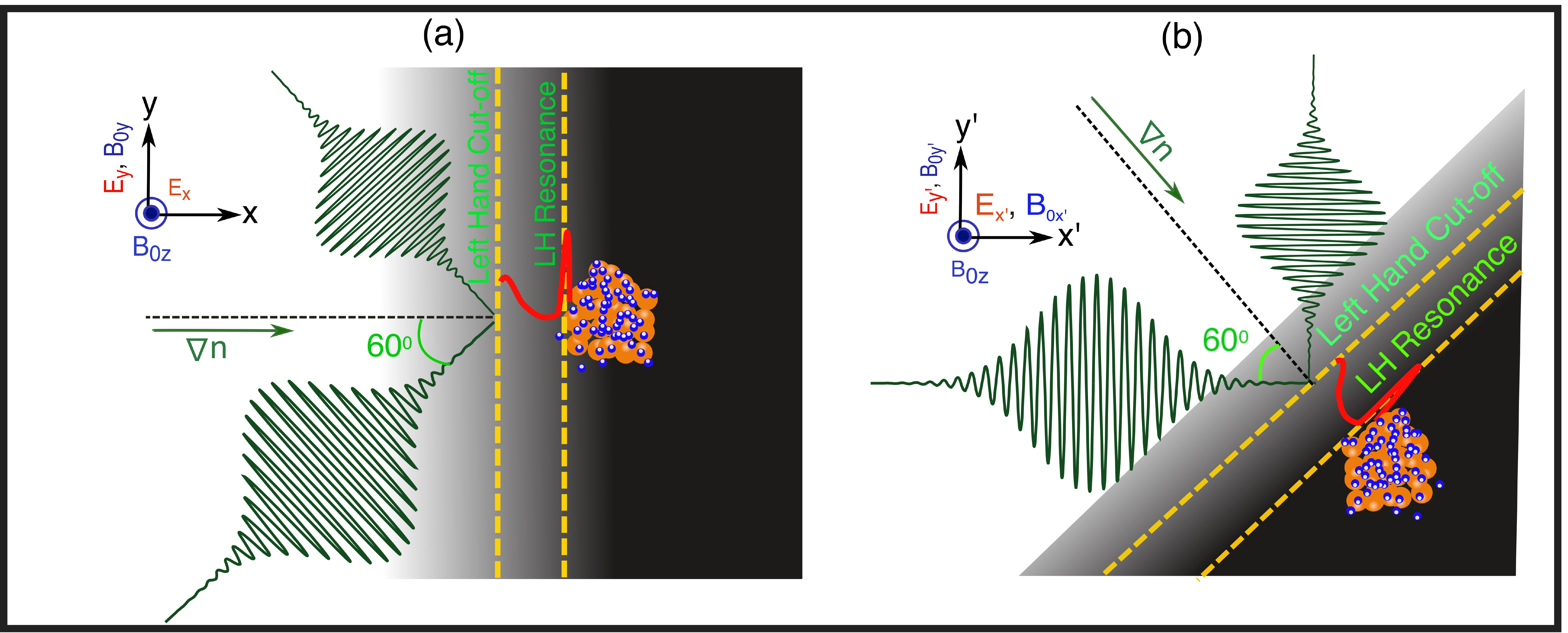} 
             \caption{ Schematics [not to scale]: (a) represents the schematics of the lower hybrid excitation scheme used in the analytical studies in x-y plane; (b) represents the lower hybrid excitation scheme for PIC simulation in the transformed gemetry $x^{\prime}-y^{\prime}$ plane where orange and blue spheres represents hot ions and electrons }  
                 \label{fig1}
         \end{figure*}

     \begin{figure*}[h!]
\center
             \includegraphics[width=\textwidth]{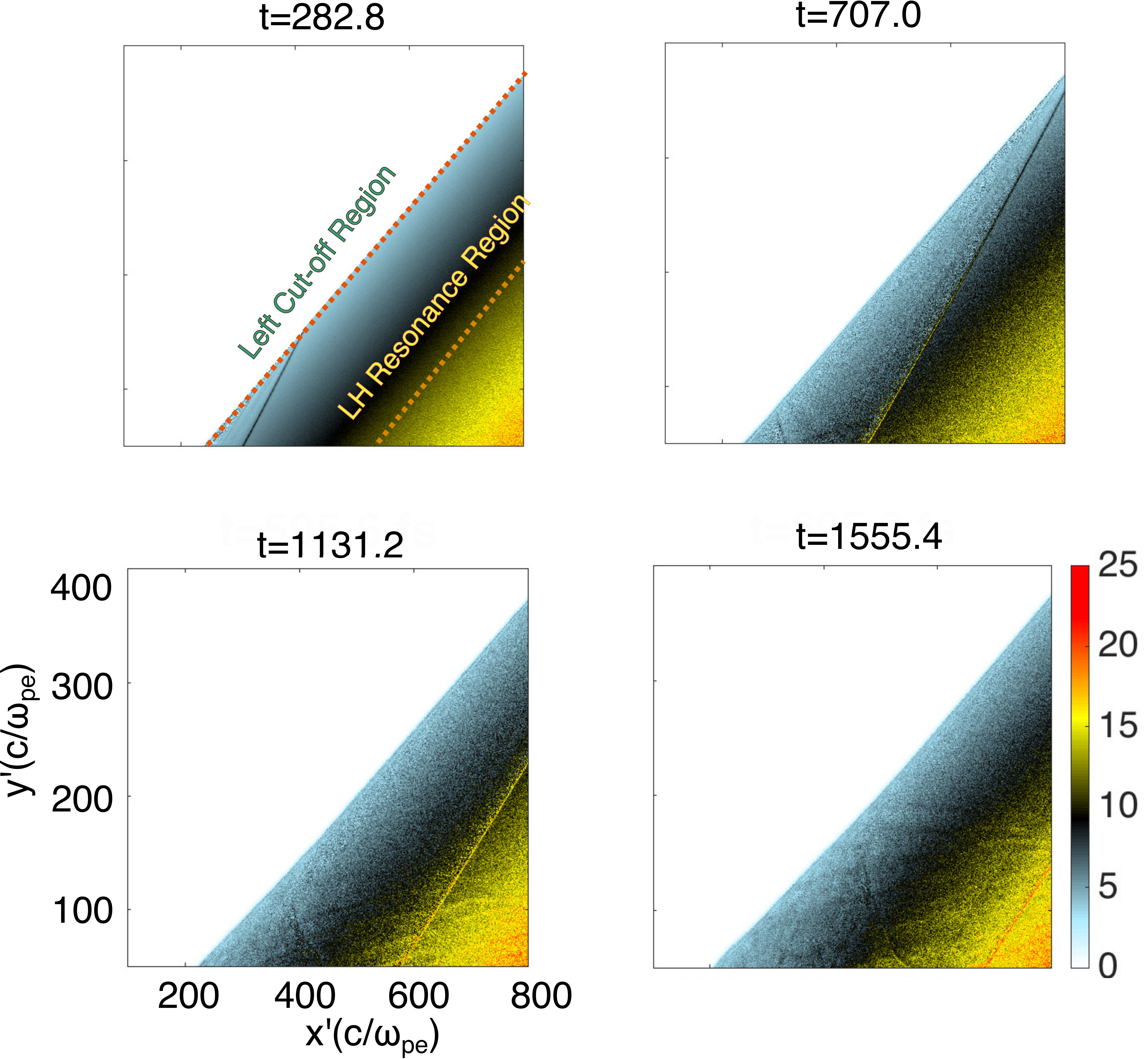}    
             \caption{Temporal evolution of the ion density which shows that the lower hybrid mode propagates beyond the resonance region }  
                 \label{fig2}
         \end{figure*} 
         
          \begin{figure*}[h!]
\center
                \includegraphics[width=\textwidth]{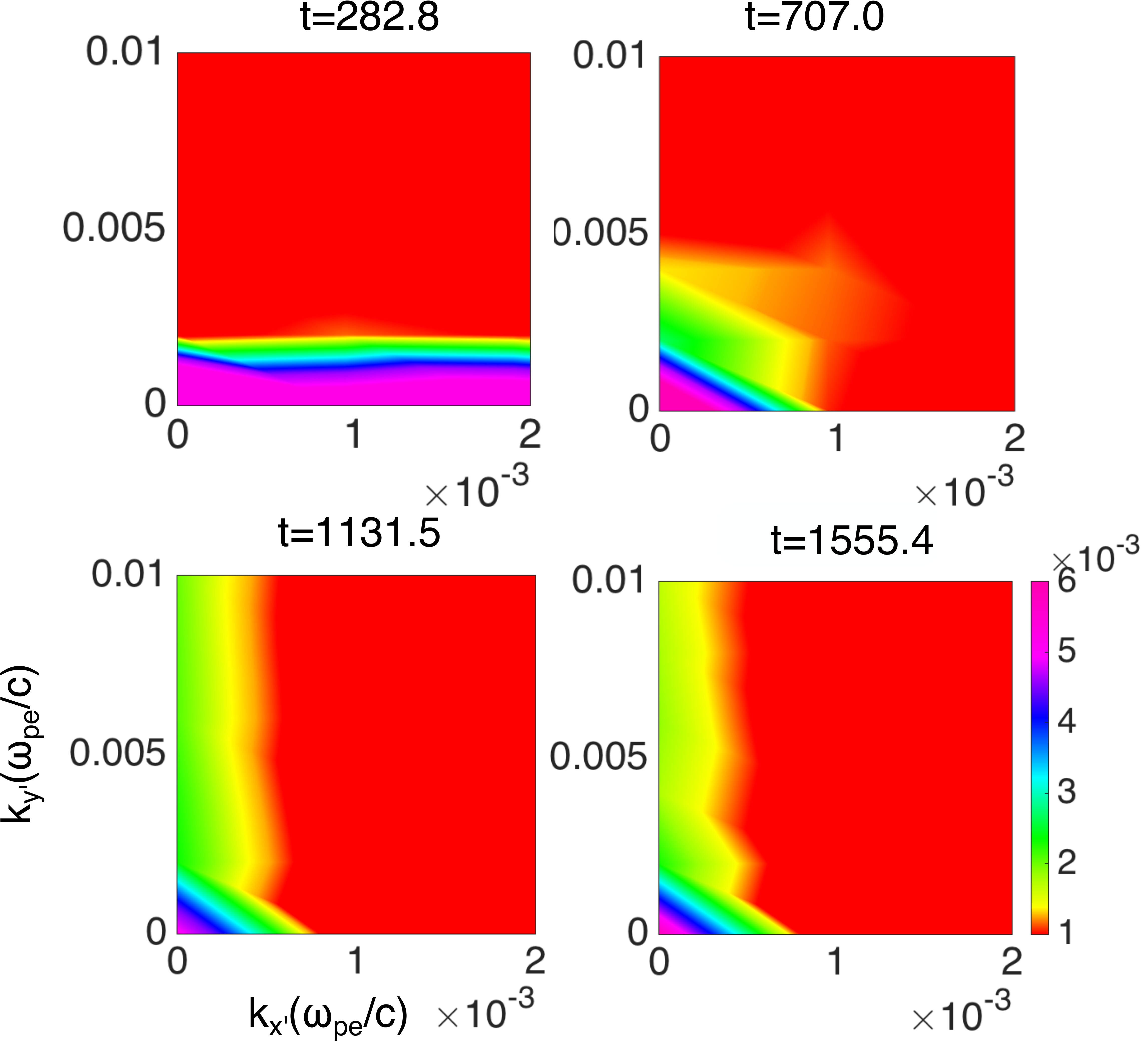} 
                
             \caption{ Temporal evolution of FFT spectra of electric field energy showing the mode conversion of X-mode into the lower hybrid mode  }

                 \label{fig3}
         \end{figure*}   
           \begin{figure*}[h!]
\center
                \includegraphics[width=\textwidth]{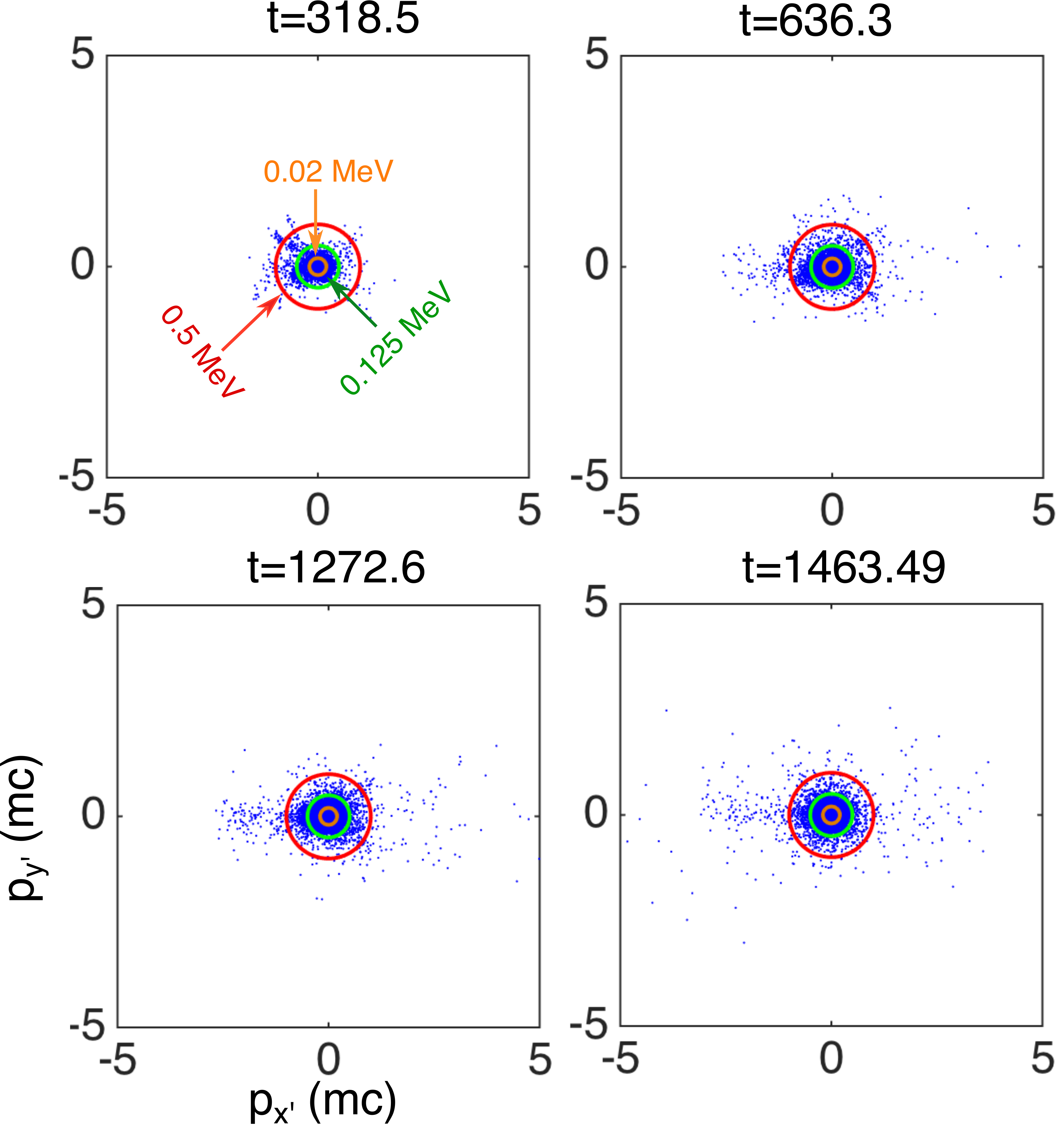} 
             \caption{Electron momentum distribution in the absence of the external magnetic field $B_{0z}$}  
                 \label{fig4}
         \end{figure*} 
         
          \begin{figure*}[h!]
\center
                \includegraphics[width=\textwidth]{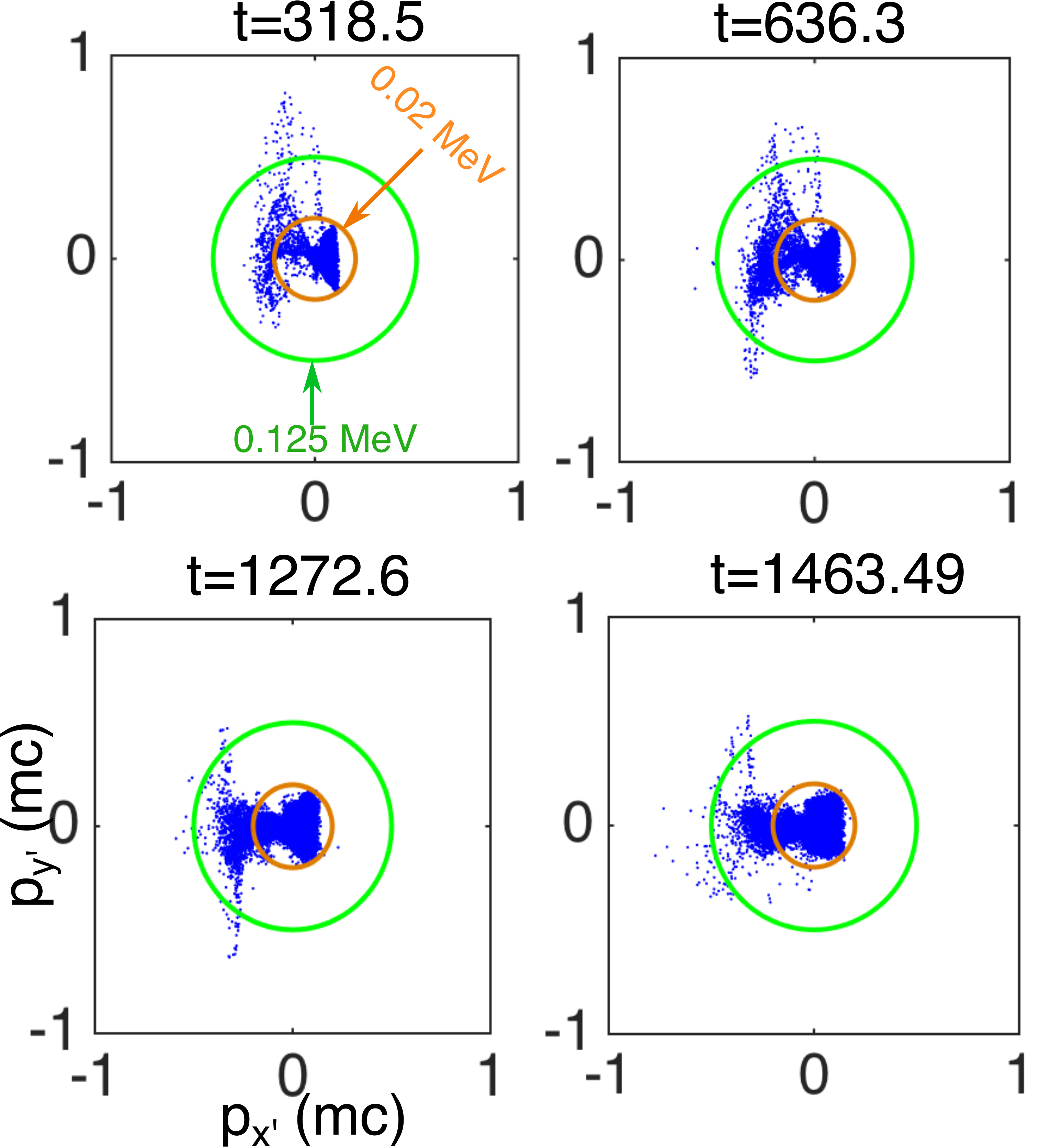} 
             \caption{Ion momentum distribution in the absence of the external magnetic field $B_{0z}$.  }

                 \label{fig5}
         \end{figure*} 
           \begin{figure*}[h!]
\center
                \includegraphics[width=\textwidth]{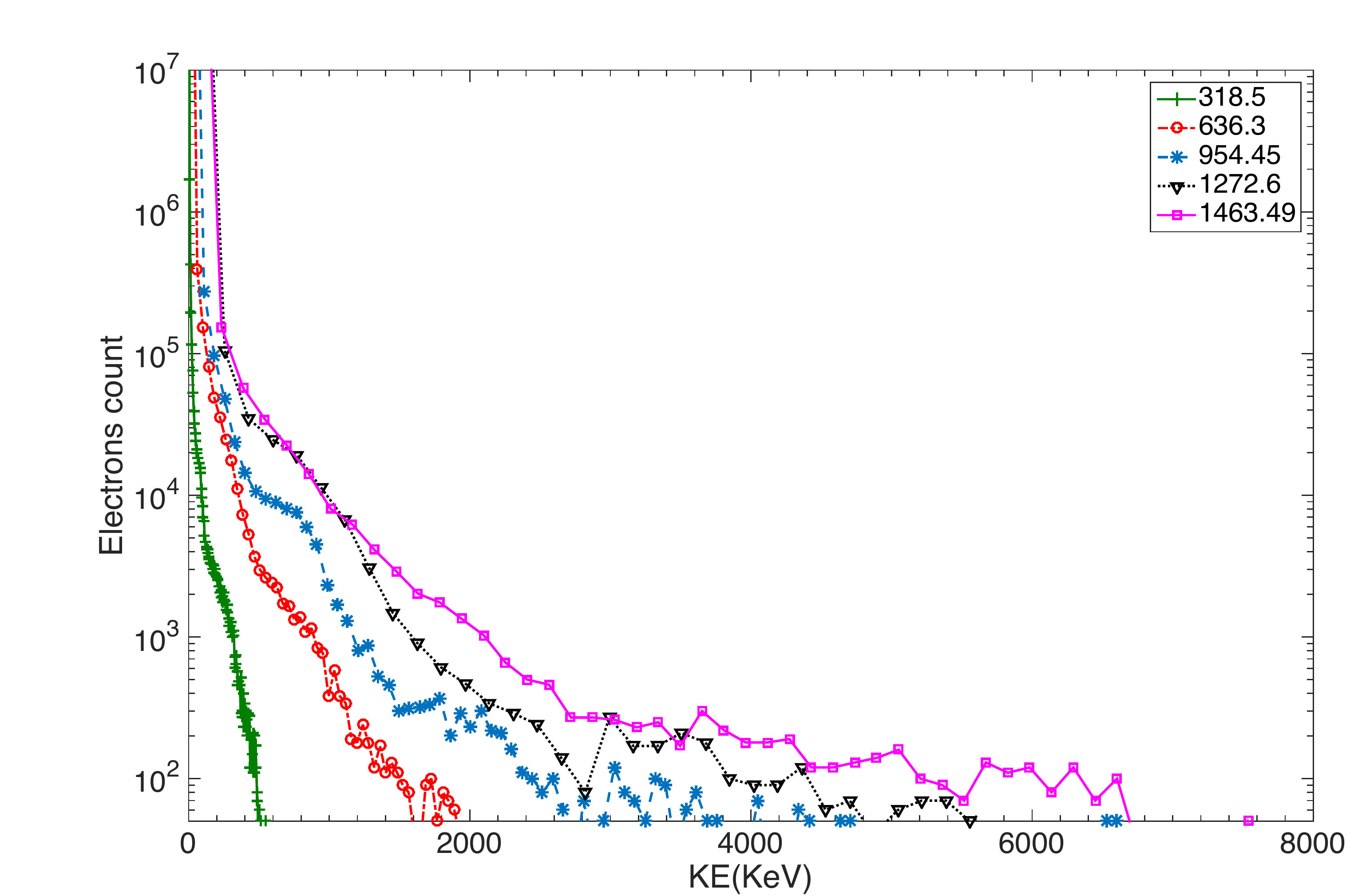} 
             \caption{ Temporal evolution of the electron energy distribution function without external magnetic field $B_{0z}$. In the absence of external magnetic field, the electron takes away almost all the absorbed energy ($\sim 8 MeV$) }  
                 \label{fig6}
         \end{figure*}
         
        \begin{figure*}[h!]
\center
                \includegraphics[width=\textwidth]{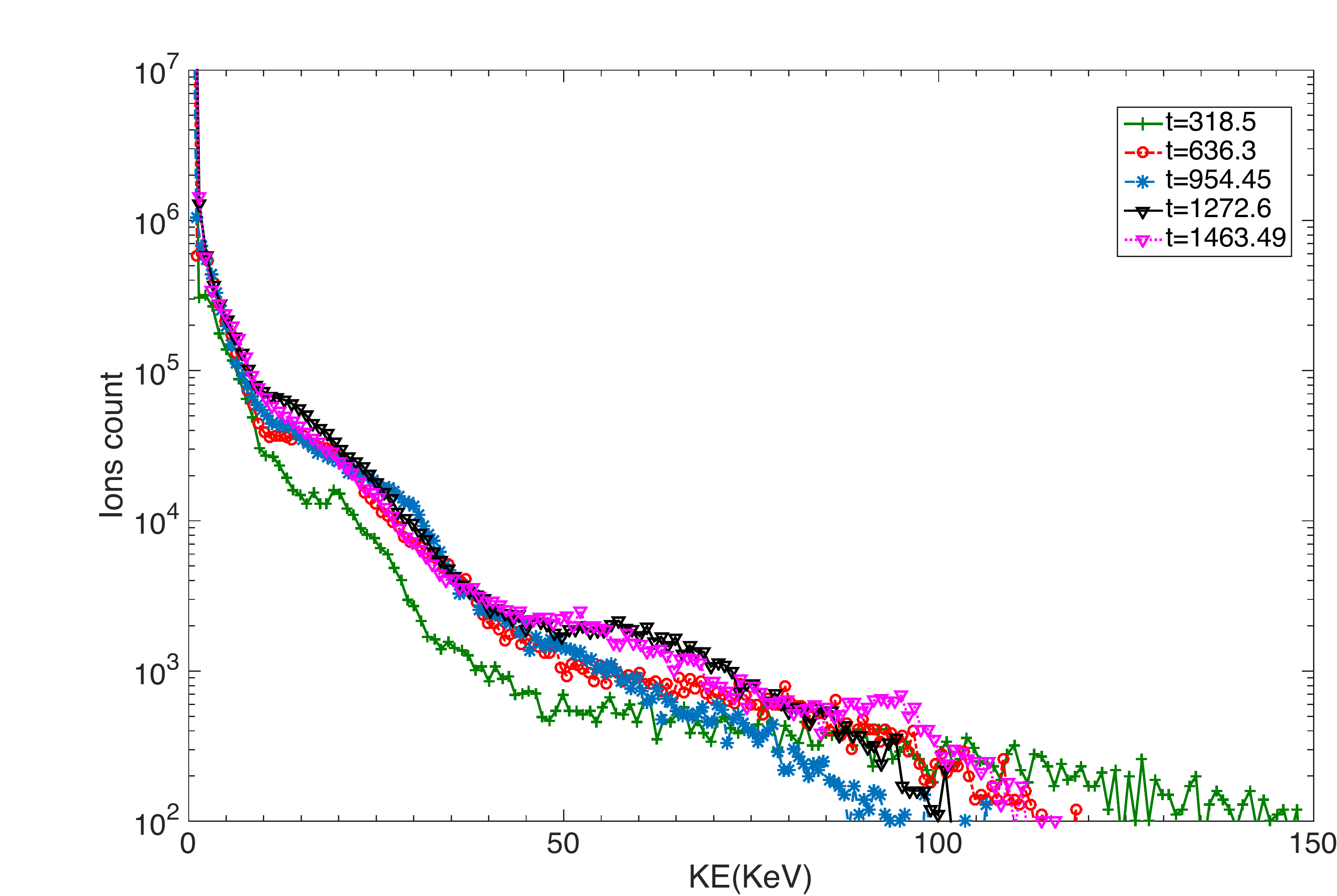}
                \caption{Temporal evolution of the ion energy distribution function without external magnetic field $B_{0z}$ showing that the ion acquire very less energy ($\sim 150 KeV$) from the incident laser  }               
                 \label{fig7}
         \end{figure*}
            \begin{figure*}[h!]
\center
                \includegraphics[width=\textwidth]{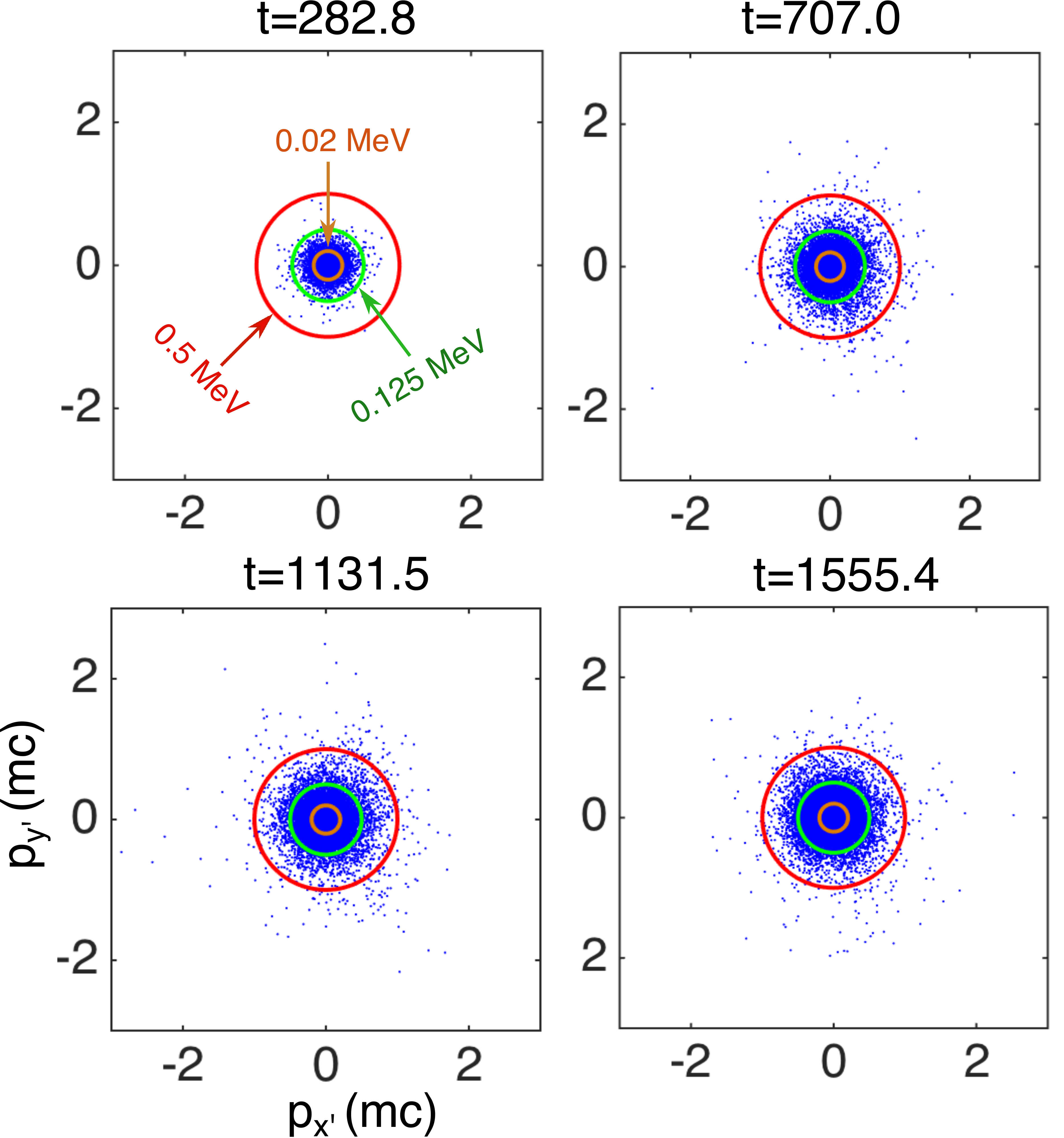} 
             \caption{Electron momentum distribution in the presence of the external magnetic field $B_{0z}$. }  
                 \label{fig8}
         \end{figure*} 
         
          \begin{figure*}[h!]
\center
                \includegraphics[width=\textwidth]{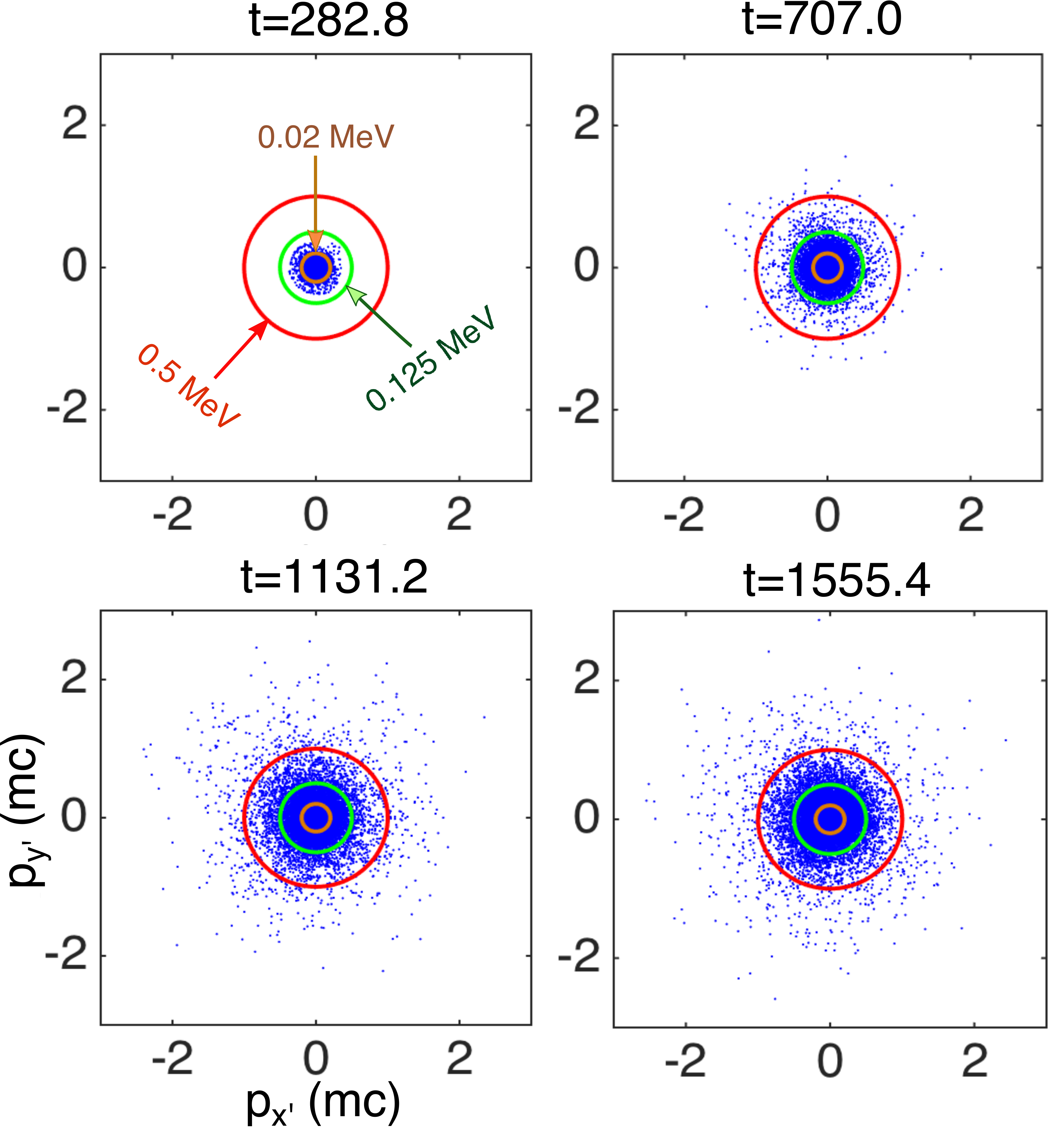} 
             \caption{  Ion momentum distribution in the presence of the external magnetic field $B_{0z}$}

                 \label{fig9}
         \end{figure*}

         \begin{figure*}[h!]
\center
                \includegraphics[width=\textwidth]{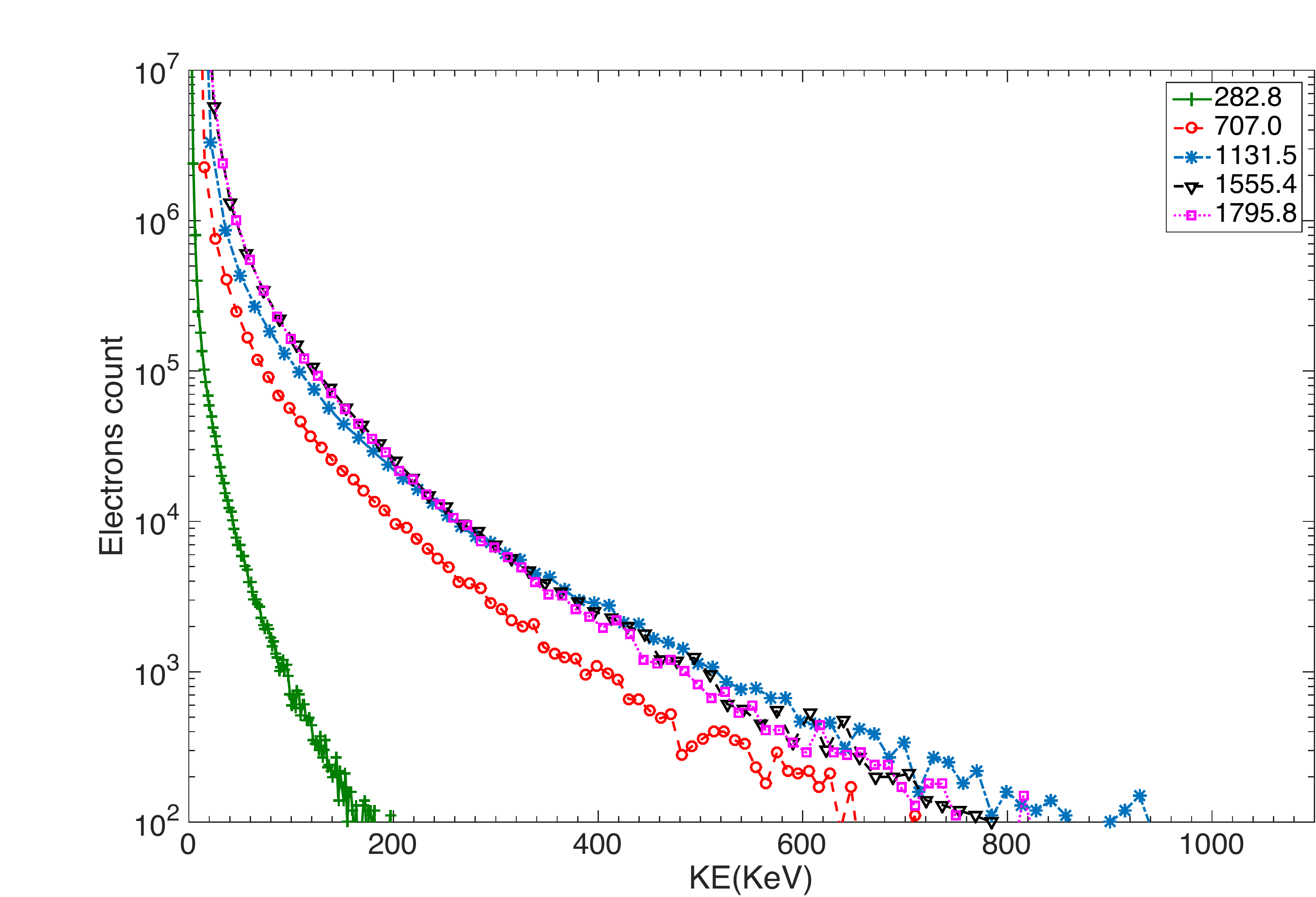} 
                \caption{ Temporal evolution of the electron energy distribution function in the presence of the external magnetic field $B_{0z}$ which shows that electrons do not takes all the fractional absorbed energy of the laser as they are binded to the external magnetic field }               
                 \label{fig10}
         \end{figure*}           
        \begin{figure*}[h!]
\center
                \includegraphics[width=\textwidth]{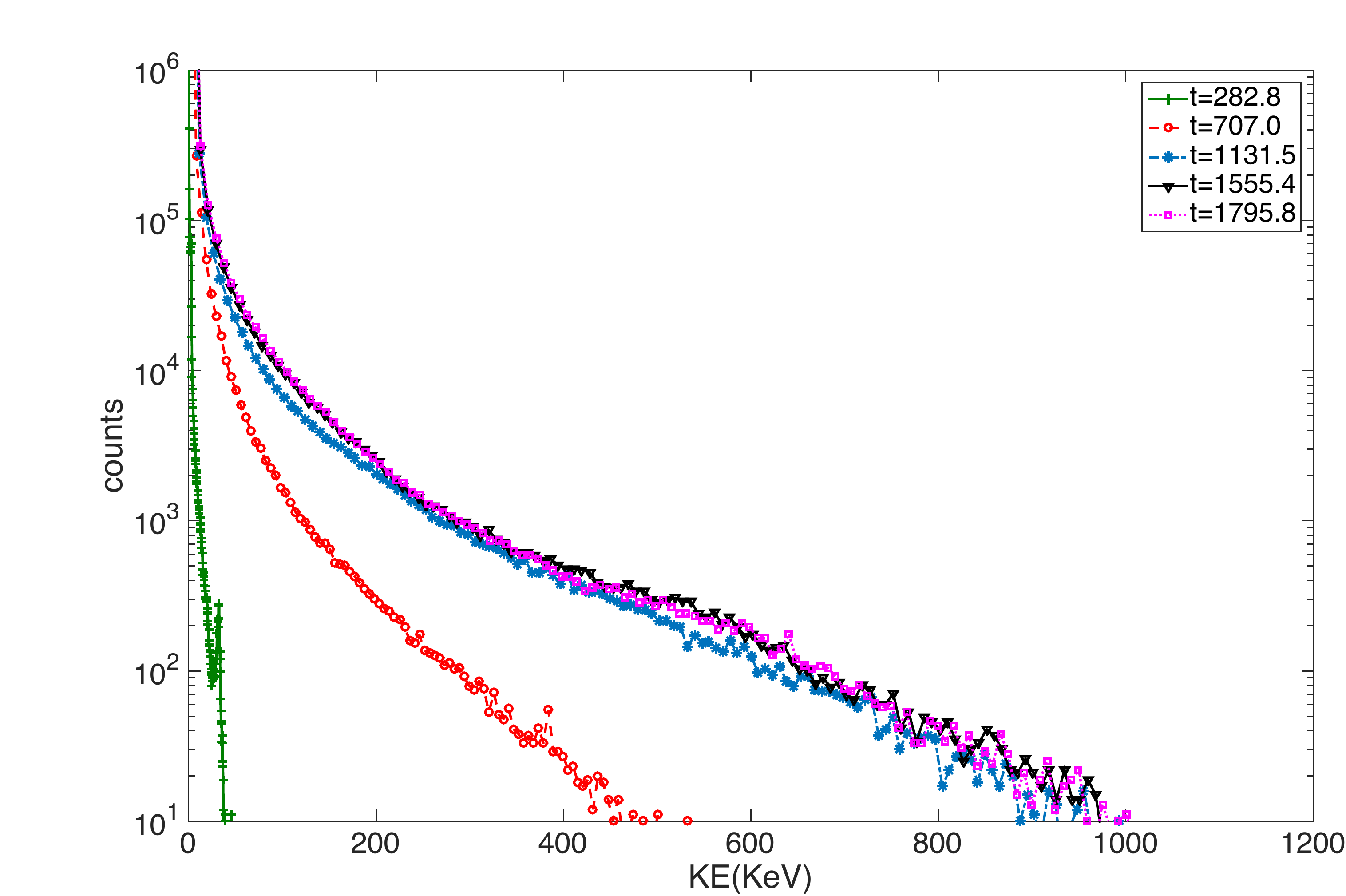} 
                \caption{  Temporal evolution of the ion energy distribution function in the presence of the external magnetic field $B_{0z}$. Here there is an equipartition of fractional absorbed energy of the incident laser between the electrons and ions which is a clear indication of lower hybrid heating }               
                 \label{fig11}
         \end{figure*}

\end{document}